\documentclass[prb,aps,twocolumn,nofootinbib]{revtex4}
\usepackage[dvips]{graphicx}
\usepackage{latexsym}
\usepackage{amsmath}
\renewcommand{\(}{\left (}
\renewcommand{\)}{\right )}

\newcommand{\vect}[1]{\boldsymbol{#1}}

\newcommand{\bea}{\begin{eqnarray}}
\newcommand{\eea}{\end{eqnarray}}

\begin{document}
\title{Charge Friedel oscillations in a Mott insulator}
\author{David F. Mross, and T. Senthil}
\affiliation{Department of Physics, Massachusetts Institute of Technology,
Cambridge, MA 02139, USA }

\begin{abstract}
When a metal undergoes a transition to an insulator it will lose its electronic Fermi surface. Interestingly in some situations a `ghost' Fermi surface of electrically neutral spin carrying fermions may survive into the insulator. Such a novel ghost Fermi surface has been proposed to underlie the properties of a few different materials but its direct detection has proven elusive. In this paper we show that the ghost Fermi surface leads to slowly decaying spatial oscillations of the electron density near impurities or other defects. These and related oscillations stem from the sharpness of the ghost Fermi surface and are direct analogs of the familiar Friedel oscillations in metals. The oscillation period contains geometric information about the shape of the ghost Fermi surface which can be potentially exploited to detect its existence.

 \end{abstract}
\maketitle
\section{Introduction}

In recent years it has become clear that the insulating side in the vicinity of the Mott metal-insulator transition may, in some frustrated lattices, provide a realization of the long sought quantum spin liquid state. These are insulators with an odd number of electrons per unit cell and that do not order magnetically or break lattice symmetries which would double the unit cell.
Examples are the layered quasi-two dimensional organic materials\cite{kanoda0,kato} $\kappa-(ET)_2Cu_2(CN)_3$ and $EtMe_3 Sb[Pd(dmit)_2]_2$, and the three dimensional `hyperkagome' material\cite{takagi} $Na_4Ir_3O_8$.  These materials can all be  driven metallic by application of moderate pressure\cite{kanoda1,takagi1}, and hence are appropriately thought of as {\em weak} Mott insulators which are not too deep into the Mott side of the metal-insulator transition. 
Indeed quantum spin liquid behavior may well be a common fate of weak Mott insulators. In contrast, the quantum spin liquid state remains much more elusive in strong Mott insulators with geometric frustration. A notable exception is the spin-$1/2$ Kagome magnet\cite{yslee} $ZnCu_3(OH)_6Cl_2$.

A fundamental question about the Mott transition is the fate of the Fermi surface of the metal as it transitions into the Mott insulator. Theoretically the natural spin liquid state in the vicinity of the Mott transition retains a `ghost Fermi surface' of mobile neutral spin carrying fermionic excitations (dubbed spinons) even as the charge of the electron gets pinned to the lattice to form an insulator. Such a state has been proposed to describe the {\em intermediate} temperature physics of many of the candidate spin liquid materials\cite{lesik,leesq,lawler}.  Experimentally both organic salts and $Na_4Ir_3O_8$ have gapless spin excitations with a low temperature specific heat\cite{syamashita} and spin susceptibility\cite{kanoda0,kato} resembling that of a metal even though they are insulating for electrical transport. Further recent experiments\cite{myamashita} in $EtMe_3 Sb[Pd(dmit)_2]_2$ show metallic thermal transport in this insulator, suggesting that the gapless excitations are, in fact, mobile.



How can we really tell if such a ghost Fermi surface of spinons survives in a weak Mott insulator? If the spinon Fermi surface state survives to low-$T$, Motrunich\cite{lesik1} has discussed the possibility of unusual quantum oscillation phenomena in a magnetic field to detect the Fermi surface. However in practice it is likely that the spinon Fermi surface undergoes an instability at low-$T$, possibly related to spinon pairing\cite{amp,triplet,orgz2}. Consequently  other less direct methods to detect it at intermediate $T$ are called for. Another idea\cite{norman} is to look for an oscillatory coupling between two ferromagnets separated by a spin liquid material due to RKKY spin correlations associated with the spinon Fermi surface. However unlike the standard giant magnetoresistance effect the oscillatory coupling cannot be readily detected by measuring the electrical resistance, and some further idea is required. This scheme also requires controlling the thickness of the spin liquid layer to atomic precision, and hence is challenging.

In this paper we identify new and surprising aspects of such a spin liquid Mott insulator that may potentially offer a simpler route to detecting the spinon Fermi surface. First we show that the electrical charge density in response to an external potential has Friedel oscillations at the ``$2K_\text{F}$" wavevectors that connect tangential antipodal portions of the spinon Fermi surface.  Friedel oscillations in the charge density are a well known property of metallic systems with a sharp electron Fermi surface. Remarkably these oscillations survive the transition into the spin liquid Mott insulating state. The spinons that form the Fermi surface in this spin liquid carry spin, and hence there are $2K_\text{F}$ singularities in the spin response. However spinons are electrically neutral - despite that the $2K_\text{F}$ singularities also show up in the charge response.

Fundamentally the charge Friedel oscillations stem from $2K_\text{F}$ singularities in the {\em spinon number density} response. This is a spin singlet response and should be distinguished from the {\em spin density} response which is a spin triplet. In a weak Mott insulator, short wavelength fluctuations of the spinon density will couple to the electrical charge density  and lead to the charge Friedel oscillations.

 A closely related effect is the existence of a Kohn anomaly at the $2K_\text{F}$ wavevectors in the phonon spectrum arising
 from the coupling of the phonons to the spinon density.  Another related effect will appear in Scanning Tunneling Microscopy (STM) spectra
 above the threshold bias for tunneling into the Mott insulator. Near the threshold the tunneling spectrum will show spatial modulation near impurities with wavelength set by the $2K_\text{F}$  wavevectors of the spinon Fermi surface. These related effects are  more directly measurable in experiments and can serve to detect the spinon Fermi surface.


We provide an estimate of the magnitude of the oscillation amplitude of the charge density and show that it is essentially unchanged as the material undergoes the metal-insulator transition.  However it is rather weak compared to the free Fermi gas so that its direct measurement may be even more challenging that in a normal metal.  Not surprisingly,  the oscillation magnitude decreases with increasing Mott gap,  eventually going to zero in the extreme limit of a spin model with no charge degrees of freedom. In contrast the Kohn anomaly is not similarly weakened and survives even into the strong Mott insulator limit.
Just as in a normal metal, it is therefore likely easier to detect the Friedel oscillations  through their effects on the phonon and STM spectra.

\section{ Spinon fermi surfaces and $2K_\text{F}$ singularities in a weak Mott insulator}
 For concreteness consider a one band Hubbard model at half-filling on a non-bipartite lattice such as the triangular lattice:
\begin{equation}
H = -t\sum_{<rr'>\alpha} \left(c^\dagger_{r\alpha}c_{r'\alpha} + h.c \right) + U\sum_r \left(n_r - 1\right)^2
\end{equation}
where $c_{r\alpha}$ destroys an electron with spin $\alpha = \uparrow, \downarrow$ at site $r$. $n_r = c^\dagger_rc_r$ is the electron number at site $r$. $U > 0$ is an on-site repulsion. As the ratio $g = t/U$ decreases the system undergoes a Mott transition from a Fermi liquid metal to a Mott insulator. 
We are interested in situations in which this Mott insulator is a non-magnetic spin liquid with gapless spin excitations.
The Mott transition and the spin liquid phase are conveniently discussed using the slave rotor representation of Ref. \onlinecite{florens}. We write
$c_{r\alpha} = e^{i\phi_r} f_{r\alpha}$
with $e^{i\phi_r} \equiv b_r$ a spin-$0$ charge-$e$ boson, and $f_{r\alpha}$ a spin-$1/2$ charge-$0$ fermionic spinon.   The electrical charge at a site $n_r$ is identified with the boson number and is conjugate to the boson phase $\phi_r$.
The physical electron operator is manifestly invariant under local opposite phase rotations of $b_r$ and $f_{r\alpha}$ respectively. A proper reformulation of the Hubbard model in terms of the $b_r$
and $f_{r\alpha}$ will thus include an emergent $U(1)$ gauge field.


An effective theory\cite{leesq} for the phases of the Hubbard model is given by an imaginary time path integral with action\small
\begin{align}
S & =  \int d\tau {\cal L}_b \left(b, a_\mu \right) + {\cal L}_f \left(f, a_\mu\right)\label{eqn:gauge1} \\
{\cal L}_b & =  \sum_r \frac{1}{2U}|\left(\partial_\tau + i a_o + \lambda \right) b_r|^2 - \sum_{<rr'>}t_b \left(b^*_r b_r' e^{ia_{rr'}} + c.c \right) \label{eqn:gauge2}\\
{\cal L}_f & =  \sum_r \bar{f}_r \left(\partial_\tau - \left(ia_o + \lambda\right)\right)f_r - \sum_{<rr'>} t_f \left(\bar{f}_r f_{r'} e^{-ia_{rr'}} + c.c \right)\label{eqn:gauge3}
\end{align}\normalsize
Here $a_0, a_{rr'}$ are the temporal and spatial components of the emergent $U(1)$ gauge field. The parameters $t_{b,f}$ are boson and spinon hopping amplitudes which may be estimated in a mean field approximation. $\lambda$ is adjusted to ensure that $\langle n_b - n_f \rangle = 0$.

In a mean field approximation, which ignores the gauge but not other interactions, the charge sector is mapped to a boson Hubbard model with hopping $t_b$ and boson repulsion $U$ while the spin sector is described by free fermionic spinons that form a Fermi surface.
When $t/U$ is large, the boson hopping $t_b$ dominates leading to a state with $\langle b_r \rangle \neq 0$. In terms of the original electrons this is a Fermi liquid phase. As $t/U$ is decreased, the bosons form a Mott insulating state with a charge gap. This corresponds to a spin liquid electronic Mott insulating state which retains a Fermi surface of gapless neutral spin-$1/2$ spinons. Beyond mean field,
at low energies in the Mott insulating phase the bosons may be integrated out, resulting in an effective theory of a spinon Fermi surface coupled to a $U(1)$ gauge field.


The possibility of sharp $2K_\text{F}$ singularities for spinons was addressed in an early work\cite{aliomil}. Very recently controlled calculations\cite{fsgauge} of $2K_\text{F}$ singularities in both the spinon number $n_{f,r} = f^\dagger_r f_r$ and the spin $\vec S_r = \frac{f^\dagger_r \vec \sigma f_r}{2}$ have been performed. 
Consider for instance the static density response function $\chi_f({\bf p})$ of the spinons at a wavevector ${\bf p}$.
 If ${\bf Q}$ is any $2K_\text{F}$ wavevector that connects two antipodal tangential portions of the spinon Fermi surface then $\chi_f({\bf Q}+{\bf q})$ is singular for small $|{\bf q}|$:
\begin{equation}
\chi^s_{f,{\bf Q}}({\bf q}) \equiv \chi_f({\bf Q}+{\bf q})-\chi_f({\bf Q})= |q_\parallel|^{\phi} F\left(\frac{q_{\perp}^2}{|q_\parallel|}\right),
\end{equation}
where $q_\parallel, q_\perp$ are the magnitudes of the components of ${\bf q}$ that are parallel and perpendicular to ${\bf Q}$ respectively. The exponent $\phi$ and the scaling function $F$ depend on spatial dimensionality. For instance in two dimensions and in the absence of the gauge field the exponent $\phi = 1/2$. The presence of the gauge field modifies the exponent in a manner that is described in Ref. \onlinecite{fsgauge}. In three dimensions the gauge field plays a more innocuous role and the exponent is the same as in the usual three dimensional Fermi liquid. Thus the spinon density correlations have sharp singular structure at all such $2K_\text{F}$ wavevectors of the underlying spinon Fermi surface.
Exactly the same exponent $\phi$ and scaling function $F$ also describe the $2K_\text{F}$ singularities in the spin density correlations due to an emergent symmetry of the spinon Fermi surface state\cite{fsgauge}.

\section{ Charge density correlations}
We now show that in a {\em weak} Mott insulator the universal short distance $2K_\text{F}$ oscillation in $n_{f,r}$ imprints itself on the correlations of the physical electrical charge $n_r$ as well. Further  the magnitude of the oscillations are more or less unchanged across the metal-insulator transition.  

The spinon number $n_{f,r}$ is a spin singlet operator that transforms under lattice and time reversal symmetries in the same way as the electrical charge $n_r = c^\dagger_r c_r$. Thus on general symmetry grounds we might expect that there is a linear coupling between $n_f$ and $n$ so that any structure in the $n_f$ correlator is transferred to $n$. For {\em long wavelength} external perturbations that couples to the charge density (which is the boson density), the boson system is gapped and the charge response is that of an incompressible insulator. However for {\em short wavelength} perturbations the boson density will have a non-zero response which, in turn, induces a response of the spinon density.  If the wavevector matches a $2K_\text{F}$ wavevector of the spinon Fermi surface, there will be long range Friedel oscillations of the fermion density that couple back to the physical charge density which will therefore show Friedel oscillations.

An approximate self-consistent calculation of the magnitude of this effect may be done within the slave particle mean field theory. An external potential with Fourier components $V({\bf p})$ induces some change $\delta \lambda$ in the chemical potential. In general the hopping parameters $t_b, t_f$ will also change, for a rough estimate we ignore these below. The resulting change in the boson and fermion densities may be expressed in linear response theory as
\begin{eqnarray}
\delta \langle n \rangle ({\bf p})& = & \chi_b ({\bf p}) \left( V({\bf p}) + \delta \lambda({\bf p}) \right) \\
\delta \langle n_f \rangle ({\bf p}) & = & -\chi_f ({\bf p}) \delta \lambda({\bf p})
\end{eqnarray}
Here $\chi_{b,f}$ are the density response function of the boson and fermion respectively.
The condition $ \delta \langle n \rangle ({\bf p}) = \delta \langle n_f \rangle ({\bf p})$ determines $\delta \lambda$. The charge density response function defined through $ \delta \langle n \rangle ({\bf p}) = \chi ({\bf p})  V({\bf p})$ then takes the form
\begin{equation}
\chi = \frac{\chi_b \chi_f}{\chi_b + \chi_f}
\end{equation}
This is the well-known Ioffe-Larkin composition rule which we now use to evaluate the $2K_\text{F}$ response.

Near a $2K_\text{F}$ wavevector $\bf{p} \approx \bf{Q}$, the spinon density response $\chi_f$ has the singularity discussed above. However as we move across the metal-insulator transition, though the small $q$ response of the boson change strikingly the high $q$ response  will be only weakly modified. We write the fermion density response  $\chi_f({\bf Q + q}) = \chi_{\bf f, 2K_\text{F}}^0 + \chi_{\bf f, 2K_\text{F}}^s(\bf{q})$ where $\chi_{\bf f, 2K_\text{F}}^0$ is the smooth background response at $2K_\text{F}$ and $\chi_{\bf f, 2K_\text{F}}^s(\bf{q})$ is the singular part.
 Denoting
$\chi_b({\bf Q + q}) = \chi_{b}^0$, we get for the singular part of the electrical density response
\begin{equation}
\chi^s_{\bf 2K_\text{F}} = R \chi_{ f,\bf 2K_\text{F}}^s
\end{equation}
where the reduction factor $R$ is given by
\begin{equation}
R = \frac{1}{\left(1+ \frac{\chi_{ f,\bf 2K_\text{F}}^0}{\chi_b^0}\right)^2}
\end{equation}
$R$ is the amount by which the Friedel oscillations of the electrical charge density are reduced compared to the oscillations of the spinons. As $\frac{t}{U}$ is varied across the Mott transition both $\chi_b^0$ and $\chi_{f, 2K_\text{F}}^0$ will be essentially unchanged so that the factor $R$, and hence the magnitude of the charge Friedel oscillations,  is more or less the same on either side of the transition. $R$ may be estimated within the slave rotor mean field theory coupled with a large-$N$ approximation to the quantum rotor model that describes the bosons (see Appendix \ref{app:rotor}).   Near the Mott transition, we find  $R = 0.024$. Though the magnitude of the charge Friedel oscillations is thus very small it is interesting conceptually that it is there at all. In the next section we will see how we may directly probe in experiments the $2K_\text{F}$ singularities of the spinon density response. Going  deeper into the Mott insulator decreases $R$ . The extreme limit of a spin model corresponds to $\chi_b(q) = 0$ for all $q$ and $R = 0$ as expected.

Actually as discussed above the detailed nature of the $2K_\text{F}$ singularity for the spinons is modified and even potentially enhanced\cite{fsgauge} in the Mott insulator as compared to the Fermi liquid due to the coupling to a gapless $U(1)$ gauge field. Thus we reach the remarkable conclusion that not only are there charge Friedel oscillations in the insulator but they may even be more slowly decaying than in the metal.

\section{Detection of spinon density Friedel oscillations}
The presence of the $2K_\text{F}$ singularities in the charge density response relies fundamentally on the $2K_\text{F}$ singularities in the spinon density response. We now argue that this latter singularity directly affects the phonon and STM spectra which could be more directly useful to detect the spinon Fermi surface.

\begin{figure}
\includegraphics[width=6.75cm]{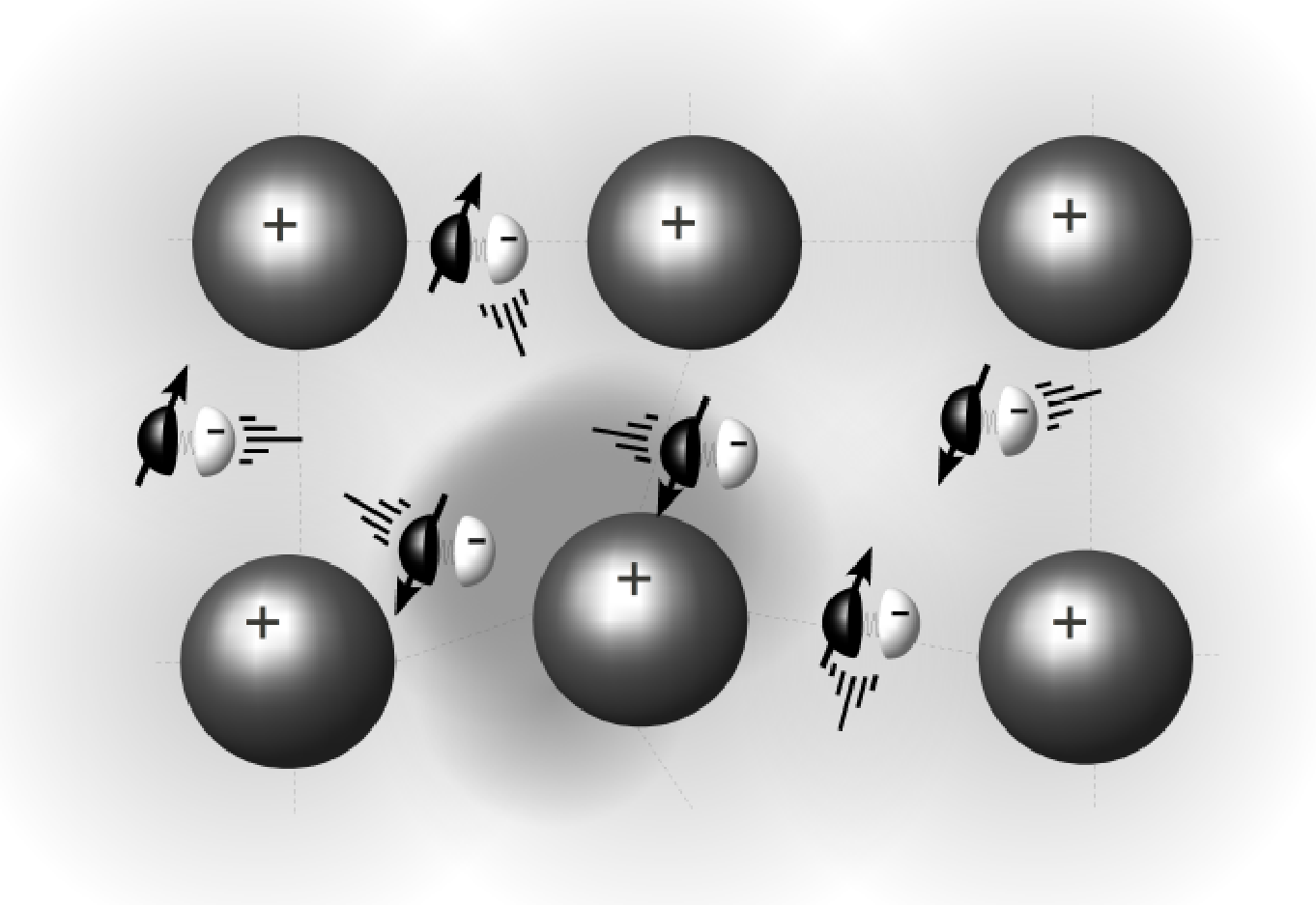}
\caption{In a regular metal, charge (represented by white half-spheres) and spin (black half-spheres with arrows) of the electrons are confined, and the electrons move to screen any displaced positively charge ions (big spheres).}
\label{fig:FL}
\end{figure}

\begin{figure}
\includegraphics[width=6.75cm]{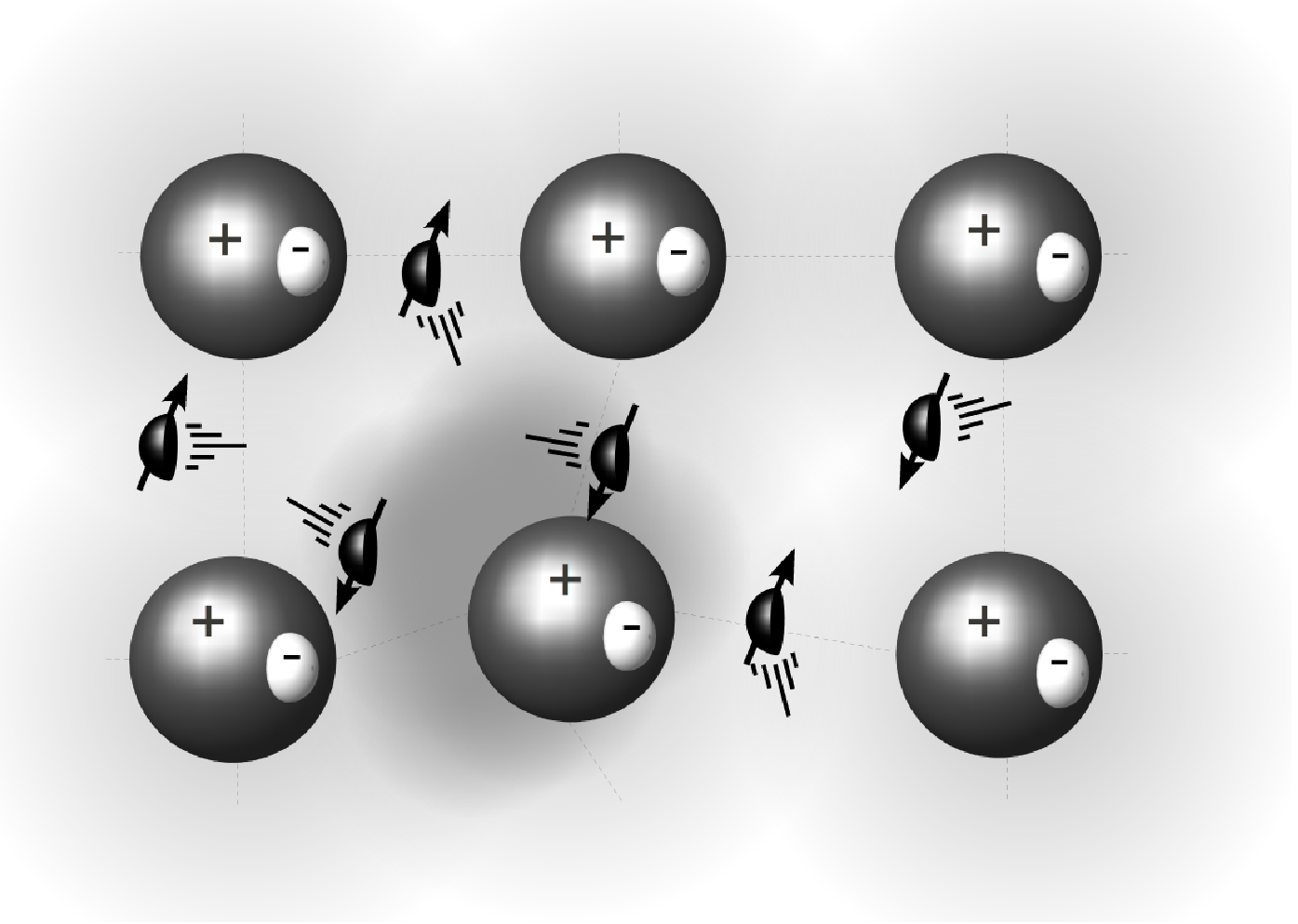}
\caption{In the spin liquid Mott insulator, the charge of the electron is bound to the ions while the spins stay mobile. The electrically neutral ion-chargon composite carry internal gauge charge however and their displacement is screened by the (oppositely) gauge-charged spinon fluid. }
\label{fig:SL}
\end{figure}

In a metal the singular $2K_\text{F}$ density fluctuations imprint themselves on the phonon spectrum through the famous Kohn anomaly. Here we show that a similar Kohn anomaly occurs in the  spin liquid Mott insulator.
In the spin liquid, the charge of the electron gets pinned to the ions that make up the underlying lattice, while the spin continues to be mobile and forms a Fermi surface. Thus the oscillating entity in a phonon mode is the electrically neutral ion-boson composite (see Figs. \ref{fig:FL},\ref{fig:SL}). However this neutral object carries gauge charge of the boson $b_r$, and hence there will be long range interactions between different ion-boson composites.  Consequently to properly describe the phonon spectrum we must take into account the screening of the emergent gauge interaction by the spinon fluid. Specifically there will be an emergent `Coulomb' interaction between the ion-boson composites that will be screened by the mobile spinons. As in the usual discussions of the Kohn anomaly the ability of the spinon fluid to screen the gauge interaction is sharply changed as the phonon wavevector passes through any $2K_\text{F}$ wavevector. Consequently there is a Kohn anomaly. This argument also makes clear that the Kohn anomaly will survive even deep into the Mott insulator so long as the system retains the spinon Fermi surface. Thus it is not weakened by the suppression of electric charge fluctuations, and will be more or less unchanged from a weakly interacting metal.

In metals Friedel-like oscillations have been imaged through Scanning Tunneling Microscopy. The most common strategy is to measure the spatial modulation of the tunneling spectrum at fixed bias. This modulation comes from the standing wave pattern caused by interference between plane wave states that are mixed by defects.
Strictly speaking these are distinct from the Friedel oscillatons but at low bias the modulation is again at the $2K_\text{F}$ wavevectors of the Fermi surface.
In the Mott insulator we are faced with the immediate problem that (at $T = 0$) there is a gap in the single electron spectrum. However if the bias exceeds the gap it will be possible to tunnel into the sample. We show below that the tunneling spectrum at a bias voltage close to the gap will acquire spatial modulation at the $2K_\text{F}$ wavevectors of the spinon Fermi surface.   The tunneling spectrum near the threshold is readily calculated within the slave rotor mean field theory.
The local tunneling density of states $N(\omega)$ is then a convolution of the boson and fermion densities of states $N_{b,f}(\omega)$:
\begin{equation}
N(\omega) = \int d \Omega N_b\left(\Omega \right) N_f\left(\omega - \Omega \right)\left(\theta(\left(\Omega\right) - \theta\left(\Omega - \omega\right)\right)
\end{equation}
In Appendix \ref{app:dos} we show that $N(\omega) = 0$ for $|\omega| < \Delta$, and increases linearly just beyond the threshold, {\em i.e}
\begin{equation}
N(\omega) = A \theta(|\omega| -\Delta)|\left(|\omega| - \Delta|\right).
\end{equation}

\small
\begin{table}

\caption{Smallest wave-vectors connecting antipodal tangential portions of the Fermi surface in the spin-liquid candidates}

 \begin{tabular}{c|c|c}\hline
$\kappa-(ET)_2Cu_2(CN)_3$ & $ZnCu_3(OH)_6Cl_2$&$Na_4Ir_3O_8$\\
\hline\hline
Triangular&Kagome&Hyperkagome\\\label{table2kf}
$0.58\frac{ \pi}{ a}\approx 0.22 \buildrel _\circ \over {\mathrm{A}}^{-1}$&$0.58\frac{ \pi}{ a}\approx 0.265 \buildrel _\circ \over {\mathrm{A}}^{-1}$&hole: $0.52\frac{\pi}{a}\approx 0.18\buildrel _\circ \over {\mathrm{A}}^{-1}$\\
&&el.: $0.62\frac{\pi}{a}\approx 0.21\buildrel _\circ \over {\mathrm{A}}^{-1}$\\
\hline
 \end{tabular}
\end{table}
\normalsize

In the presence of a defect such as a step edge or an impurity, $N_f(\omega; {\bf x})$ oscillates as a function of spatial location ${\bf x}$ at the $2K_\text{F}$ wavevectors of the spinon Fermi surface for low frequency $\omega \rightarrow 0$ . $N_b(\omega)$ will also have some ${\bf x}$ dependence but will not oscillate at the same wavevectors. Near the threshold the tunneling electron creates a boson with energy close to $\Delta$ and  a fermion close to zero energy.  As shown in Appendix \ref{app:dos} the coefficient $A$ will show spatial modulation at the $2K_\text{F}$ wavevectors.



\begin{figure}
\includegraphics[width=6cm]{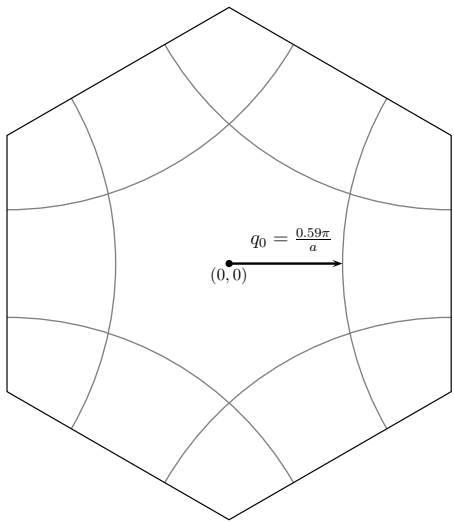}
\label{2kfsurf}
\caption{$2k_F$ vectors for the triangular lattice at half filling in the 1st Brillouin zone.}
\end{figure}

We now comment  on the observability of these effects in the candidate spin materials in hand, namely the quasi-two dimensional 
$\kappa-(ET)_2Cu_2(CN)_3$ and $EtMe_3 Sb[Pd(dmit)_2]_2$, and the three dimensional $Na_4Ir_3O_8$.
In all three materials determination of the phonon spectrum (to look for the Kohn anomaly) through neutron scattering is challenging but may possibly be done through inelastic X-ray scattering techniques. In the organic materials the spinon Fermi temperature is estimated to be $\sim 250 K$. Consequently the Kohn anomaly will only be visible at much lower temperature. Furthermore due to the layered structure it will be independent of the component of momentum perpendicular to the layers. Some representative wavevectors along high symmetry directions are given in Table \ref{table2kf}, and the surface of all $2K_\text{F}$ wavevectors in the first Brillouin zone is shown in Fig. \ref{2kfsurf} assuming a perfect triangular lattice. The hyper-kagome material $Na_4Ir_3O_8$ has a rather complicated band structure in its metallic state which will be inherited by the spinons in the proposed spin liquid Mott insulator. Some representative $2K_\text{F}$ wavevectors were calculated for this material in Ref. \onlinecite{norman}, and we list them here for completeness in Table \ref{table2kf}.

Previous STM studies\cite{stmorg}  of the $\kappa-ET$ organics have focused on a superconducting material. STM studies of the insulating spin liquid will be useful both in determining the single particle gap and in detecting the possible spinon Fermi surface.
In this context we note that in $\kappa-(ET)_2Cu_2(CN)_3$ the charge gap as measured by optical conductivity and dc transport is small (in the 15 -100 meV range) though it is hard to extract a precise number.  This range  should be readily accessible in STM experiments.  

Finally we comment briefly on strong Mott insulators with negligible virtual charge fluctuations but which nevertheless are in quantum spin liquid phases. This is the situation, for instance, in the Kagome material $ZnCu_3(OH)_6Cl_2$. Such systems are well described by simple spin-only models with dominant nearest neighbor exchange. Spin liquids with spinon Fermi surfaces may emerge in these spin-only models as well. Indeed such a spin liquid has been proposed for $ZnCu_3(OH)_6Cl_2$ in Ref. \onlinecite{marston}. What is the interpretation of the $2K_\text{F}$ singularities in the spinon number $n_f$ in that case? As this is a spin singlet this singular $2K_\text{F}$ response signals slow power correlations of incommensurate valence bond solid order. Such correlations have been found in a spin-1/2 system on a triangular strip \cite{sheng}. The arguments presented above show that
  phonons of the lattice will couple to these incommensurate valence bond fluctuations (similarly to the familiar one dimensional gapless spin liquids), and there will again be a Kohn anomaly in the phonon spectrum at the $2K_\text{F}$ wavevectors. Detecting such an anomaly will then be a non-trivial proof of the proposal that a spinon Fermi surface exists in this material.

We thank J. Hoffman, P.A. Lee, and Young Lee for useful discussions. TS was supported by NSF Grant DMR-0705255.
\appendix
\section{ Mean field theory in the large $N$ limit}\label{app:rotor}
In the mean field theory, the fermion action is gaussian but the boson action is still interacting. We generalize the $O(2)$ model a model of $N$ complex bosons $\mathbf{ b} = (b_1,\ldots,b_N)$ which we solve by expanding in powers of $1/N$ and extrapolating to $N=1$. The mean-field boson action is

\begin{align}
{\cal L}_b &= \sum_r \frac{1}{2U}\left|\left(\partial_\tau +\sum_iA^i_0\underline g^i +  \mu \right) \mathbf{ b}_r\right|^2 \\
& \ \ - \sum_{<rr'>}t_b \left(\vect{b}^*_r\cdot \vect b_{r'}  + c.c \right)+i \lambda_r(|\vect b_r|^2-N),
\end{align}
where $\underline g^i$ are the $2 N^2+\mathcal{O}(N)$ generators of $O(2N)$ and $\lambda$ is a
 Lagrange multiplier field, that is integrated over in the path integral. The \emph{external}
 gauge field $A^i_0$ should not be confused
with the \emph{emergent} gauge field $a_0$ in Eq. (\ref{eqn:gauge1}). For large $N$ we can replace $\frac{\Delta^2}{2 U} \equiv i \lambda$ by its saddle-point value, which for $g<g_c$ is given by

\begin{equation}
1 =  \int d\bar\omega \sum_{\vec k \in \text{BZ}} \frac{1}{t_b\epsilon_{\vec k}+\frac{1}{2U} \omega^2+\frac{\Delta^2}{2 U}},
\end{equation}
where $\epsilon_{\vec k}=6-\sum_{i} e^{i \vec a_i \cdot \vec k}$, with $\vec a_i$ all vectors connecting
a lattice site to its nearest neighbors. At half filling on the triangular lattice $t_b \approx 0.33 t$ and
by setting $\Delta=0$ we find $g_c \approx 0.39$. 
The boson-susceptibility $\chi_b =\sum_i \chi_b^i=\chi_b^{\text{dia}}+\chi_b^{\text{para}}$ where
the diamagnetic and paramagnetic contributions are given by
\small
\begin{eqnarray}
&&\chi_b^{\text{para}}(\vec q)=-\frac{2 N^2}{(U t_b)^2}\int d\omega\sum_{\vec k \in BZ} \frac{\omega^2}{\epsilon_{\vec k}+\frac{\omega^2+\Delta^2}{2U t_b}}\frac{1}{\epsilon_{\vec k+\vec q}+\frac{\omega^2+\Delta^2}{2U t_b}}\nonumber\\
&&\chi_b^{\text{dia}}=\frac{2 N^2}{U}=-\chi_b^{\text{para}}(0).
\end{eqnarray}\normalsize
The fermion susceptibility is given by the usual expression
\begin{equation}
\chi_f(\vec q)=N_s\int d\bar\omega \sum_{\vec k \in \text{BZ}} \frac{1}{-i\omega +t_f\epsilon_{\vec k}-\mu} \frac{1}{-i\omega +t_f\epsilon_{\vec k+\vec q}-\mu},
\end{equation}
with $N_s$ the number of spin species.
For $N=1$, $N_s=2$, $\Delta=0$, and $\vec q \equiv q_0\hat x$, we find $\chi_b (\vec q_0)\approx \frac{0.22}{U} , \chi_f (\vec q_0)\approx \frac{1.2}{U}$, so that
\begin{equation}
R(\vec q_0)=\left(1+\frac{\chi_f^0(\vec q_0)}{\chi_b(\vec q_0)}\right)^{-2}\approx 0.024
\end{equation}.
\section{Tunneling d.o.s.}\label{app:dos}
The electron Green function in real space is given by
\begin{align}
G_e(\tau,\vec r)&=  G_f(\tau,\vec r)G_b(\tau,\vec r),
\end{align}
so that
\begin{align}
N(\omega)& \equiv -\frac{1}{\pi}\text{Im} \int d^2 \bar k G^{ret}_e(\omega,\vec k)\\
& =\int d \Omega N_b\left(\Omega \right) N_f\left(\omega - \Omega \right)\left(\theta\left(\Omega\right) - \theta\left(\Omega - \omega\right)\right),\label{eqs:dosapp}
\end{align}
where the Heaviside-functions $\theta$ are a consequence of the pole structure of the retarded Green functions. First, note that the fermion d.o.s. is constant at the Fermi energy, i.e.
\begin{align}
\lim \limits_{\omega\rightarrow 0}N_f(\omega,\vec x) =N_f(\epsilon_F,\vec x) +\mathcal{O}\(\frac{\omega}{\epsilon_F}\).
\end{align}
Next consider the boson d.o.s.
\begin{align}
N_b(\omega)=\int d^2\bar k \delta\(\frac{\omega^2-\Delta^2}{2U}-t_b \epsilon_k\).
\end{align}
Clearly $N_b(\omega)=0$ for $|\omega|<\Delta$ and for $\omega-\Delta \ll \Delta$ we can expand the dispersion at the bottom of the band $\epsilon_k\approx \frac{3}{2}k^2$ to obtain
\begin{align}
\lim \limits_{\omega \rightarrow \Delta }N_b(\omega)=\frac{1}{2t}\theta(\omega-\Delta)+\mathcal{O}\(\frac{\omega}{\Delta}\).
\end{align}
Plugging both these expression into Eq. (\ref{eqs:dosapp}), we obtain
\begin{align}
N(\omega)&=\frac{1}{2t}\theta(\omega-\Delta)\int_0^{\omega-\Delta} d \Omega  N_f\left(\omega-\Delta - \Omega \right)\\
&=(\omega-\Delta)\theta(\omega-\Delta)\frac{N_f(\epsilon_F,\vec x)}{2t}.
\end{align}


\end{document}